\documentclass[conference]{IEEEtran}
\IEEEoverridecommandlockouts
\usepackage{cite}
\usepackage{amsmath,amssymb,amsfonts}
\usepackage{algorithmic}
\usepackage{graphicx}
\usepackage{textcomp}
\usepackage{url}
\usepackage{xcolor}
\usepackage{float}
\usepackage{caption}
\usepackage{makecell}
 \usepackage{balance}
\def\BibTeX{{\rm B\kern-.05em{\sc i\kern-.025em b}\kern-.08em
    T\kern-.1667em\lower.7ex\hbox{E}\kern-.125emX}}
\begin{document}

\title{Nowcasting of Aviation Radiation Using Geospace Environment Properties: A Machine Learning Approach}

\author{\IEEEauthorblockN{1\textsuperscript{st} Sanjib K C}
\IEEEauthorblockA{\textit{Department of Physics and Astronomy} \\
\textit{Georgia State University}\\
Atlanta, USA \\
skc3@gsu.edu}
\and
\IEEEauthorblockN{2\textsuperscript{nd} Viacheslav M Sadykov}
\IEEEauthorblockA{\textit{Department of Physics and Astronomy} \\
\textit{Georgia State University}\\
Atlanta, USA\\
vsadykov@gsu.edu}
\and
\IEEEauthorblockN{3\textsuperscript{rd} Dustin Kempton}
\IEEEauthorblockA{\textit{Department of computer Science} \\
\textit{Georgia State University}\\
Atlanta, USA\\
dkempton1@gsu.edu}
}

\maketitle

\begin{abstract}
Radiation exposure at aviation altitudes presents significant health risks to aircrews due to the cumulative effects of ionizing radiation. Physics-based models estimate radiation levels based on geophysical and atmospheric parameters, but often struggle to capture the highly dynamic and complex nature of the radiation environment, limiting their real-time predictive capabilities. To address this challenge, we investigate machine learning (ML) methods to enhance real-time radiation nowcasting. Leveraging newly compiled, ML-ready datasets~--—publicly available at \cite{url-rdpmldataset}~--— we train supervised models capable of capturing both linear and non-linear relationships between Geospace conditions and atmospheric radiation levels. Our experiments demonstrate that the XGBoost model achieves approximately 10 percent improvement in prediction accuracy over the considered physics-based model. Furthermore, feature importance analysis reveals that certain Geospace properties, specifically solar polar fields, solar wind properties, and neutron monitor data, are impacting the nowcast of the radiation levels at flight altitudes. These findings suggest meaningful physical relationships between the near-Earth space environment and atmospheric radiation, and highlight the potential of ML-based approaches for operational space weather applications.
\end{abstract}

\begin{IEEEkeywords}
Space Weather, Radiation Environment, Machine Learning
\end{IEEEkeywords}

\section{Introduction}
Studies of radiation at aviation altitudes are essential for the safety of aviation crews, flight passengers, and aviation electronics, as well as for understanding the terrestrial radiation environment. Radiation environment at aviation altitudes is primarily determined by ionizing radiation from Galactic Cosmic Rays (GCRs) and Solar Energetic Particle events (SEPs) \cite{friedberg2003aircrews, Tobiska2016ARMAS,straume2016ground}. The origin of GCRs, a primary source of aviation radiation that always exists, is external to the solar system. Contrary to it, SEPs are bursts of energetic protons and heavy ions originating during solar flares or coronal mass ejections \cite{Tobiska2015ARMAS,Tobiska2016ARMAS}. Recent studies show that there might be a third radiation source, relativistic energetic particles (REP), that precipitate from the Van Allen radiation belts \cite{Tobiska2018ARMAS}. Combined, these sources determine the aviation radiation environment.

The International Commission on Radiological Protection (ICRP) recognizes aircrew as radiation-exposed workers. They also recommend an effective dose limit of 20\,mSv per year averaged over 5 years (totaling 100\,mSv in 5 years) for radiation workers. However, for the general public, the recommended limit is 1\,mSv per year \cite{valentin2007international}. Nowadays, raising concerns about health and safety at aviation altitudes has become increasingly important with the rise in commercial and research flights at aviation \cite{Tobiska2016ARMAS,sadykov2025operational}. Moreover,  several aerospace missions are being conducted by organizations like Space Environment Technologies (SET) using instruments such as the Automated Radiation Measurements for Aerospace Safety (ARMAS) dosimeter, which is flown aboard aircraft to measure real-time dose rates \cite{tobiska2019armas}. Given more than 12 years in operation, the ARMAS measurement series currently constitutes, possibly, the richest data sets for the aviation radiation environment \cite{Sadykov2021RDP}.

\begin{figure}
        \centering
        \includegraphics[width=1.00\columnwidth]{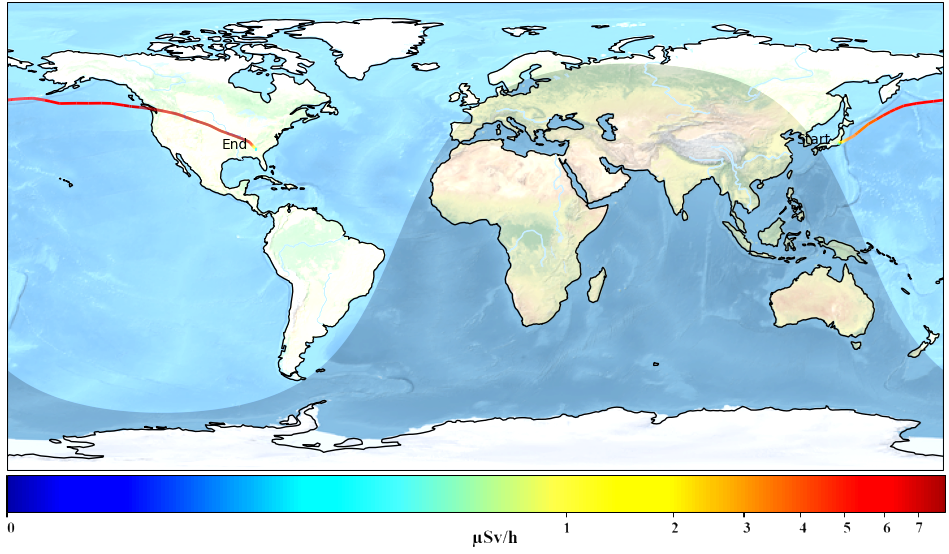}
        \caption{Effective dose rate without shielding along the Tokyo–Atlanta flight trajectory on June 7, 2025, calculated from NAIRAS simulation using the Run-on-Request (RoR) services of the Community Coordinated Modeling Center.}
        \label{fig:nairas_flightexample}
        \vspace{-0.5cm}
\end{figure}

The problem of nowcasting and forecasting the aviation radiation environment is very important, since the information can be used for flight rerouting and a more accurate accounting of the cumulative radiation doses received. Various physics-based and statistical models, such as the Nowcast of Atmospheric Ionizing Radiation for Aerospace Safety (NAIRAS) \cite{mertens2009development,Mertens2013,Mertens2023, Mertens2024, mertens2025NAIRAS}, CARI-7 \cite{copeland2021cari7}, and others, have been developed to correctly estimate the radiation exposure at the upper atmosphere \cite{phoenix2024characterization}. For example, Figure~\ref{fig:nairas_flightexample} illustrates the nowcast of the radiation for the flight DL294 from Tokyo to Atlanta that occurred on June 7, 2025, obtained using the NAIRAS-v3 model deployed at the NASA Community Coordinated Modeling Center \cite{url-ccmc-nairasv3}. Nevertheless, the complex behavior of and the non-linear interactions in the Geospace environment (i.e., the solar magnetic activity cycle, the Earth's magnetosphere, solar magnetic transients such as coronal mass ejections, etc.) continue to challenge the physics-driven models.

To counter the limitations of existing physics-based models and given the rich set of ARMAS measurements, a promising solution is Machine Learning (ML) that is intrinsically capable of modeling non-trivial dependencies in the data, and may serve as a promising substitute. For easier monitoring and access of the data for ML studies, a team of researchers (including contributors to this paper) have developed the Radiation Data Portal \cite{Sadykov2021RDP}, an interactive platform that combined ARMAS observations with Geospace variables such as geomagnetic latitude, interplanetary magnetic field (IMF) properties, solar wind conditions, solar polar field strength, neutron monitor data, and other relevant parameters. The most recent version of the portal also features the ML-ready data set for radiation environment forecasting \cite{url-rdpmldataset}. It provides the possibilities to predict the radiation environment using both static (point-in-time) and time-series measurements of the environment properties.

\section{Methodology and Data}

\subsection{Data Description}

This study utilizes the machine learning (ML)-ready dataset available from the latest (Aug. 2025) version of the \emph{Radiation Data Portal (RDP, \cite{Sadykov2021RDP}, \cite{url-rdpdmlab})}. The data set augments in-flight dosimetry measurements from the ARMAS program, with supporting Geospace environment parameters. These supporting parameters include, but are not limited to, energetic proton, electron, and X-ray fluxes from the Geostationary Operational Environmental Satellites (GOES), geomagnetic indices (Kp, Ap, and Dst), global solar activity parameters (such as daily sunspot number, 10.7\,cm radio flux, and solar polar magnetic fields), measurements from ground-based neutron monitors, measurements of the solar wind at the L1 point, etc. Our dataset spans from 2013 to 2023 and contains 47 input features plus one target variable (ARMAS radiation dose rate, in $\mu$Sv/h). The feature set largely mirrors that used in the NAIRAS-v3 model, ensuring comparability with a well-established physics-based baseline. Additionally, our dataset includes several supplementary parameters such as sunspot number, electron fluxes in channels $\geq2$\,MeV, and spatial parameters (barometric altitude, GPS latitude and longitude). These additional features provide richer environmental and spatial context to enhance the model’s predictive capability. For a detailed description of all features and their physical significance, please see \cite{sadykov2025mlradiation}.

\subsection{Data Preprocessing}

The RDP ML-ready dataset was cleaned and synchronized by removing unrealistic radiation measurement values (sampling ARMAS measurements only between barometric altitudes of 8 km - 15.5 km), temporal alignment, missing data handling, and removal of duplicate device measurements. The full dataset was partitioned into three subsets (hereafter partition1, partition2, partition3). Six rotating train-validation-test splits were therefore constructed by designating one subset as training data, another as validation, and the third as testing in all permutations. During the subset construction, it was ensured that the data samples from each unique flight are fully incorporated into a single partition \cite{sadykov2025mlradiation}, making sure there is no temporal or event leakage between the subsets.\\

\noindent Further preprocessing included:

\begin{itemize}
    \item \textbf{Feature Selection:} Non-predictive columns such as \texttt{Datetime}, \texttt{NAIRASV2}, \texttt{NAIRASV3}, and \texttt{Vehicle\_ID} were removed. For training the Least Absolute Shrinkage and Selection Operator (LASSO) model \cite{tibshirani1996regression}, only the absolute values of `Latitude ' and `Geomagnetic latitude' were used due to the linear assumptions of the model and the known physics linking latitude to radiation intensity.
    \item \textbf{Standardization:} Since LASSO requires standardized inputs for numerical stability, each feature \(x\) was normalized as
    \[
    x' = \frac{x - \mu_x}{\sigma_x},
    \]
    where \(\mu_x\) and \(\sigma_x\) are the mean and standard deviation computed from the training set.
\end{itemize}

\subsection{Machine Learning Models and Motivation}

We investigated supervised linear and tree-based algorithms to explore the feasibility of radiation nowcasting using classic ML approaches and to gain physical insight via feature analysis. The physics-based NAIRAS-v3 model \cite{Mertens2023, Mertens2024, mertens2025NAIRAS} was employed as the primary baseline, providing a reference grounded in established domain knowledge. LASSO regression was chosen as the baseline ML model, due to its built-in feature selection capability through regularization of L1, promoting sparsity and interpretability \cite{tibshirani1996regression, tibshirani2011regression}.



To capture potential nonlinear relationships between the inputs and target, Random Forest \cite{breiman2001random} and XGBoost \cite{chen2016xgboost} were also tested. These ensemble methods are widely recognized for high performance on tabular data and have shown predictive power in space weather applications \cite{ali2024predicting, Sadykov2025ionkinetic}.

\subsection{Implementation Details}

LASSO and Random Forest ML models were implemented with the \texttt{scikit-learn} \cite{pedregosa2011scikit}, Python library (version~1.7.0), and XGBoost was implemented with the \texttt{XGBoost} Python library (version~3.0.2) \cite{chen2016xgboost}. All experiments were run with model training, evaluation, and hyperparameter optimization performed on each of the six rotating data splits.

\subsection{Hyperparameter Tuning}

Hyperparameter tuning was conducted via an exhaustive brute-force grid search over predefined parameter grids for each model, on each train-validation split. For each split, multiple combinations of hyperparameters were evaluated to minimize validation root mean squared error (RMSE). This manual grid search approach allowed precise control over parameter combinations and ensured the best parameters were selected per split.

\begin{itemize}
    \item \textbf{LASSO:} Tuned the regularization strength \(\alpha\) controlling the trade-off between sparsity and fit quality, by searching for over 1000 equally spaced values in the range $\alpha \in [0.01, 1.0]$ as the optimum alpha values $\alpha$ are in the second decimal places.
    
    \item \textbf{Random Forest:} A manual grid search was performed over combinations of key hyperparameters to identify the best model configuration. The following ranges were explored: \texttt{n\_estimators} (50, 100, 300), \texttt{max\_depth} (3, 6, 10), \texttt{min\_samples\_split} (2, 5, 10), \texttt{min\_samples\_leaf} (1, 2, 4), \texttt{max\_features} ('auto', 'sqrt'), \texttt{bootstrap} (True, False), and \texttt{criterion} ('squared\_error', 'absolute\_error'). Models were trained for each parameter combination, and the best configuration was selected based on validation performance.

    \item \textbf{XGBoost:} Hyperparameters were manually tuned using an exhaustive grid search implemented with nested loops. The following ranges were explored: \texttt{n\_estimators} (50, 100, 300), \texttt{max\_depth} (3, 6, 10), \texttt{learning\_rate} (0.01, 0.1, 0.2), \texttt{subsample} (0.6, 0.8, 1.0), \texttt{colsample\_bytree} (0.6, 1.0), \texttt{reg\_alpha} (0, 0.1), and \texttt{reg\_lambda} (1, 5, 10). The best configuration was selected based on validation performance.

\end{itemize}

\subsection{Statistical Comparison}

Model performance was evaluated using root mean squared error (RMSE) and the coefficient of determination (R$^{2}$) on training, validation, and test datasets. Statistical comparisons of machine learning models against the physics-based NAIRAS-v3 baseline were conducted using these metrics to provide a quantitative assessment of predictive accuracy and explained variance.

\subsection{Feature Importances}

Feature importance was assessed separately for each model type as follows:

\begin{itemize}
    \item \textbf{LASSO:} Due to LASSO’s L1 regularization, many coefficients shrink to zero, effectively performing feature selection. To identify robustly important features across the six data splits, the following criteria were applied:
    \begin{itemize}
        \item A feature must have a non-zero coefficient in at least 3 out of 6 splits (i.e., at least 50\% of splits) with the same directionality.
        \item The mean absolute coefficient across splits must exceed 0.0001.
    \end{itemize}
    Features satisfying these criteria were ranked by the absolute value of their mean coefficient, with both positive and negative effects preserved.
    
    \item \textbf{Random Forest and XGBoost:} Feature importance values were obtained directly from the respective algorithms. Features with importance below 1\% were excluded, and no split consistency threshold was applied.
\end{itemize}

This approach ensures only features with consistent and significant contributions across multiple splits are considered important, enhancing the robustness and interpretability of feature relevance.

\section{Nowcasting of radiation environment}

Understanding and evaluating the models and their feature importance analysis is critical to interpreting the Physics behind the model. The model evaluation was performed specifically using the RMSE and $R^2$ metrics. 
For feature importance analysis, a minimum threshold of 1\% (based on impurity decrease) 
was applied for all tree-based algorithms. For \textsc{LASSO}, feature selection was instead 
based on three criteria: an absolute mean coefficient threshold of $0.0001$, appearance in at 
least $3$ cross-validation splits, and a minimum directional consistency of $50\%$.

\begin{figure*}
    \centering
    \includegraphics[width=1.4\columnwidth]{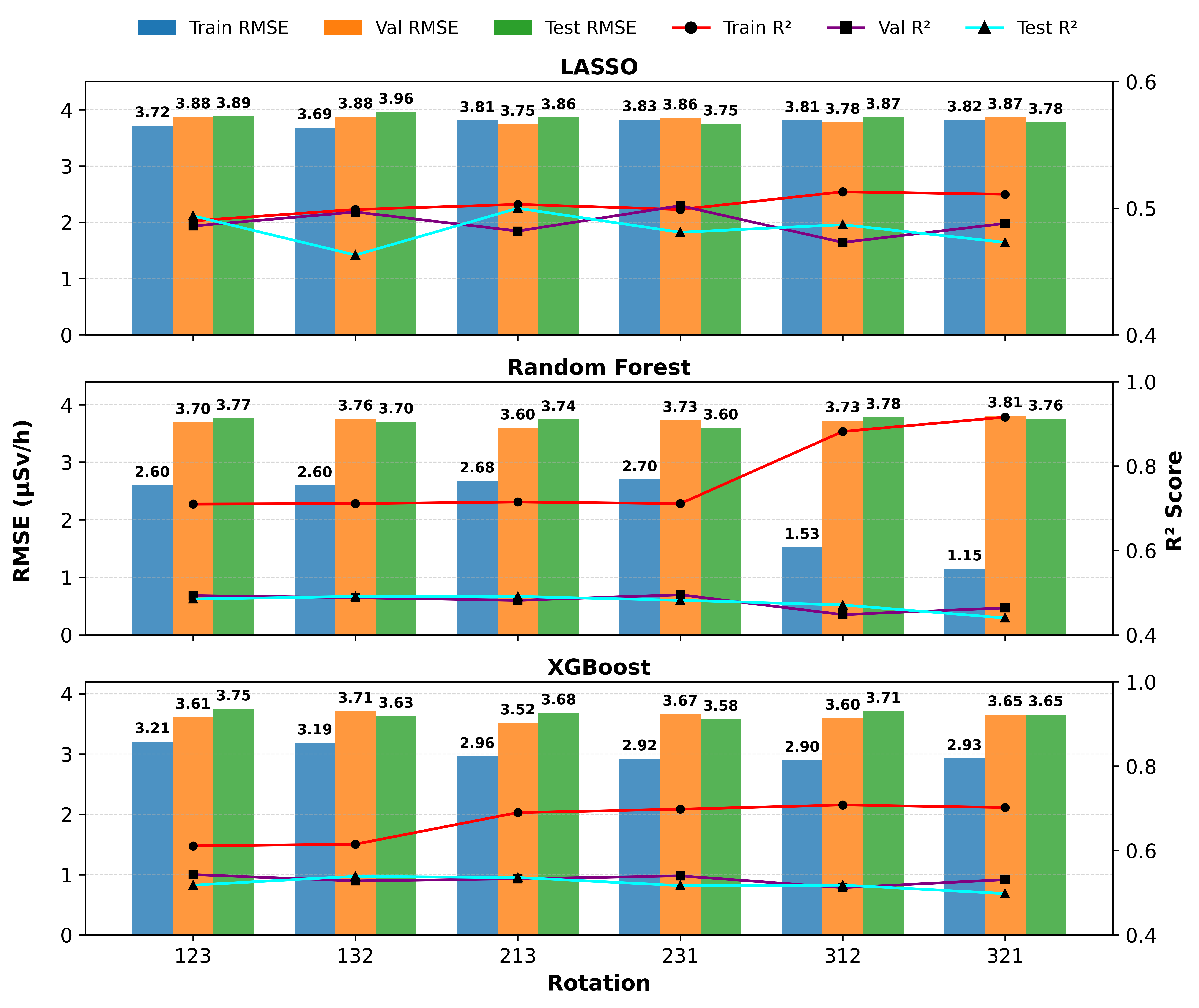}
    \caption{Comparison of the Root Mean Squared Error (RMSE) in radiation prediction for different train-calibration-test partition combinations. Note: '123' means partition 1 is train, partition 2 is validation, and partition 3 is test, and likewise.}
     \label{fig:nairas}
\end{figure*}

Physics-based NAIRAS-v3 model predictions can be directly compared against corresponding ARMAS measurements. The average RMSE from 3 partitions is 4.045$\mu$Sv/h and $R^2$ score is 0.430. Following these scores as a baseline comparison, we performed nowcasting on linear (LASSO) and tree-based algorithms (Random Forest and XGBoost) across all six train-calibration-test combinations from 3 data sets. The summary of the scores is presented in Figure~\ref{fig:nairas}, and the model average scores across the train-validation-test splits are presented in Table~\ref{tab:model_avg_metrics}.

\begin{table*}[htbp]
\centering
\resizebox{\textwidth}{!}{%
\begin{tabular}{|c|c|c|c|c|c|c|c|}
\hline
\textbf{Model} & \textbf{Train RMSE} & \textbf{Train R\textsuperscript{2}} & \textbf{Validation RMSE} & \textbf{Validation R\textsuperscript{2}} & \textbf{Test RMSE} & \textbf{Test R\textsuperscript{2}} & \makecell{\textbf{Improvement} \\ \textbf{over NAIRAS-v3 (\%)}} \\
\hline
LASSO         & 3.780 $\pm$ 0.061 & 0.503 $\pm$ 0.009 & 3.835 $\pm$ 0.056 & 0.488 $\pm$ 0.010 & 3.853 $\pm$ 0.077 & 0.483 $\pm$ 0.014 & 4.76\% \\
Random Forest & 2.209 $\pm$ 0.687 & 0.814 $\pm$ 0.097 & 3.720 $\pm$ 0.070 & 0.518 $\pm$ 0.019 & 3.725 $\pm$ 0.067 & 0.517 $\pm$ 0.019 & 7.92\% \\
XGBoost       & 3.020 $\pm$ 0.140 & 0.681 $\pm$ 0.045 & 3.630 $\pm$ 0.067 & 0.541 $\pm$ 0.011 & 3.671 $\pm$ 0.061 & 0.531 $\pm$ 0.015 & 9.26\% \\
\hline
NAIRAS        & -- & -- & -- & -- & 4.045 $\pm$ 0.087 & 0.430 $\pm$ 0.015 & -- \\
\hline
\end{tabular}%
}
\caption{Model Performance and Percentage Improvement in Test RMSE ($\mu$Sv/h) Compared to NAIRAS v3 RMSE ($\mu$Sv/h). Here, the standard deviations are computed over six train-validation-test partitions.}
\label{tab:model_avg_metrics}
\vspace{-0.5cm}
\end{table*}

\begin{figure}
    \centering
    \includegraphics[width=\columnwidth]{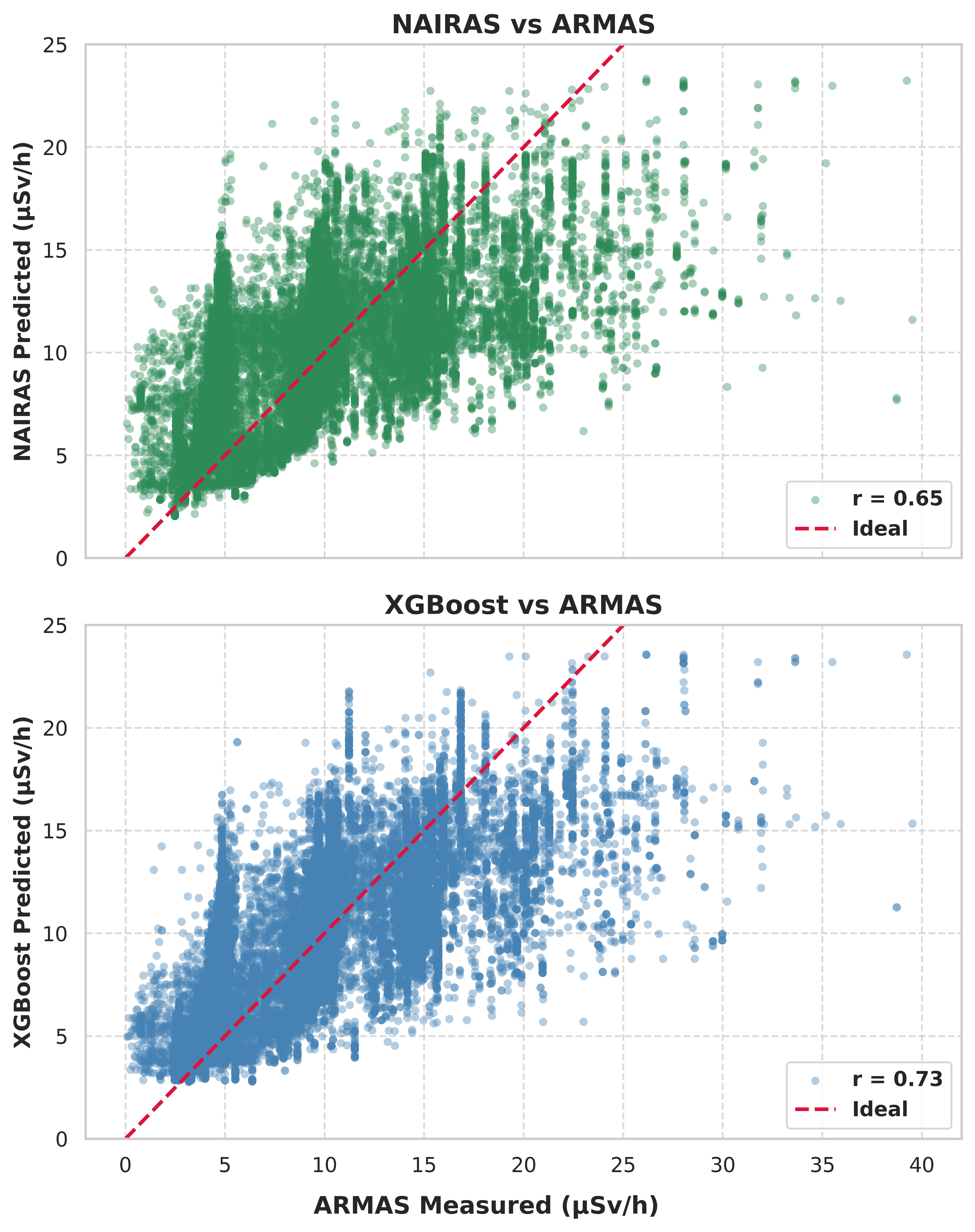}
    \caption{Measured radiation dose rates VS NAIRAS-v3 (top) and nowcasted XGBoost model (bottom) for partition 1 (split-231).
Both the NAIRAS-v3 and XGB nowcasts demonstrate the poor performance for the ‘tail’ of the distribution (measurements above $\sim$15 µSv/h).
}
     \label{fig:scatter}
    \vspace{-0.5cm}
\end{figure}

As illustrated in Figure~\ref{fig:nairas}, the RMSE scores in all partitions are below 4\,$\mu$Sv/h for all test subsets in any train-validation-test split considered for all three models. Along with the average scores summarized in Table~\ref{tab:model_avg_metrics}, this demonstrates an improvement against the NAIRAS-v3 average RMSE of 4.045\,$\mu$Sv/h across 3 partitions. The scores are mostly consistent in all the splits and corresponding train-validation-test, validating the reliability of partitioning (described in \cite{sadykov2025mlradiation}). Some peculiarities of the scores for each type of machine learning model are discussed below.

Despite being a relatively simple linear regression model, LASSO still outperforms the NAIRAS-v3 physics-based model by $\sim$5\% on average (RMSE=3.853$\mu$Sv/h, see Table~\ref{tab:model_avg_metrics}). For the LASSO regression model, no significant overfitting is seen in any split, as evident in Figure~\ref{fig:nairas}: the performances on the train-validation-test data sets are comparable.

The slightly better nowcasting of the radiation environment (RMSE=3.725$\mu$Sv/h) can be achieved for the Random Forest model. The tuned Random Forest consistently outperformed the baseline physics model across all data splits, showing improved predictive accuracy and robustness. However, in partition `321`, significant overfitting was observed. The model showed very high accuracy on the training set but a noticeable drop in validation/test performance, indicating limited generalization due to overfitting.

In the case of the tuned XGBoost model, the results are consistent in all partitions, and there are no signs of strong overfitting.  The corresponding RMSE averaged over six train-validation-split tests reaches RMSE=3.671$\mu$Sv/h, with the minimum RMSE of 3.583$\mu$Sv/h for the `231' split. This model has demonstrated the best performance among the three considered. 

The R$^{2}$ values (also presented in Figure~\ref{fig:nairas} and Table~\ref{tab:model_avg_metrics}) are consistent with the RMSE measures: the lower RMSEs correspond to the higher R$^{2}$ values. Overall, the performance of the ML models is better with respect to the corresponding physics-based NAIRAS-v3, even for the case of the static data set (based on the point-in-time environment properties) and relatively simple models considered in this work, which overall holds promise for the radiation environment nowcasting with machine learning.\\
Figure~\ref{fig:scatter} shows the scatter plot for the physics-based NAIRAS-v3 Vs ARMAS (top; green) and XGBoost Vs ARMAS (bottom; blue) plots. Both NAIRAS-v3 and XGB nowcasts demonstrate poor performance for the 'tail' of the distribution. This behavior could appear for several reasons, including the data imbalance issue due to a low number of measurements with the extreme $\gtrsim$15\,$\mu{}Sv/h$ dose rates or physical effects not captured by both models. We are planning to consider this behavior in the following works. Nevertheless, the overall agreement with ARMAS measurements differs between the models: the Pearson correlation coefficient (r) for NAIRAS-v3 Vs ARMAS is 0.65, whereas for XGB Vs ARMAS it is notably higher at 0.73, underscoring the stronger predictive capability of XGB compared to NAIRAS-v3.

\section{Feature importance analysis}

Studying the importance of various features propagating into the models can not only help to eliminate `noisy' or non-informative features, but also provide insights into what properties of the Geospace environment are important for the radiation at aviation altitudes. The latter, correspondingly, can provide ideas about the physical dependences, including those that are not captured by the current physics-based models. All three machine learning approaches considered in this work have a natural way of estimating feature importances (using linear regression coefficients for LASSO, and impurity decrease in tree-based Random Forest and XGBoost).

\subsubsection{LASSO Feature importance}
\begin{figure*}
    \centering
    \includegraphics[width=1.3\columnwidth]{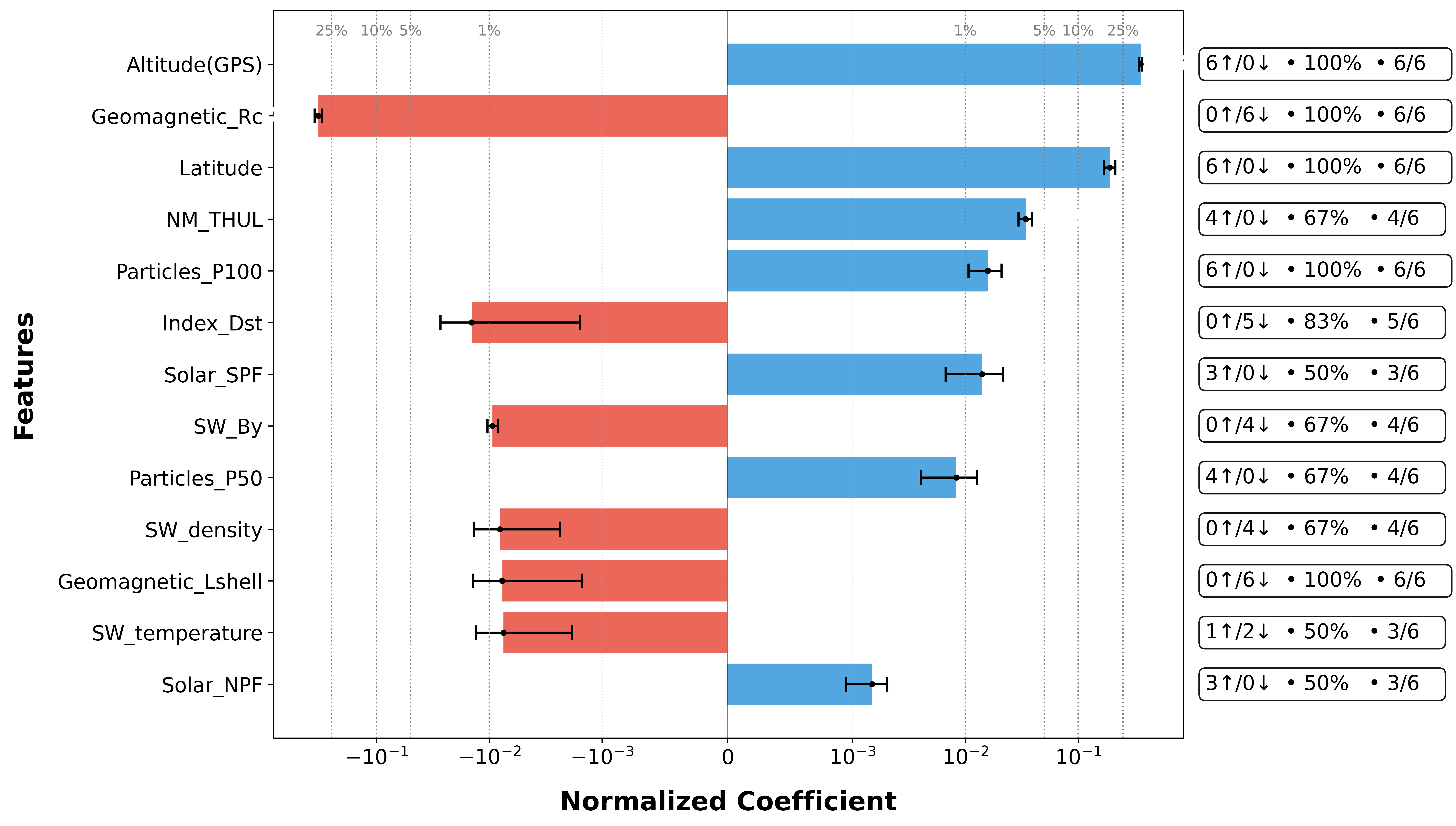}
    \caption{Feature importance according to the LASSO model. The vertical lines show 1\%, 5\%, 10\%, and 25\% normalized importance.}
    \label{fig:lassofeatureimp}
\end{figure*}

Feature importance for the LASSO was assessed through the coefficients. Figure~\ref{fig:lassofeatureimp} shows the average feature importance for the model. Feature importances were calculated for all splits, and the average importance for each feature was taken, preserving the sign. In figure~\ref{fig:lassofeatureimp}, only features contributing 1\% or more are shown. In addition, the thresholding for the minimum appearance of 3 or more was done to take the features that consistently participate in the nowcast. To ensure comparability with the tree-based model feature importances, the average coefficients were also normalized. Each horizontal bar in the figure shows the normalized mean coefficient \(\tilde{\beta}_j\), with error bars representing the confidence interval (CI) calculated as
\[
\tilde{\beta}_j \pm \tilde{\sigma}_j,
\]
where
\[
\tilde{\beta}_j = \frac{\overline{\beta}_j}{\sum_{k=1}^p |\overline{\beta}_k|} \quad \text{and} \quad \tilde{\sigma}_j = \frac{\mathrm{SD}(\overline{\beta}_j)}{\sum_{k=1}^p |\overline{\beta}_k|}.
\]
Here, \(\overline{\beta}_j\) is the mean coefficient estimate for feature \(j\) across splits or runs, and \(\mathrm{SD}(\overline{\beta}_j)\) is the standard deviation of these mean estimates. The CI illustrates the uncertainty in the estimated effect size of each feature.

\subsection{Feature Importance Visualization for tree-based models}  
\begin{figure*}
        \centering
        \includegraphics[width=1.3\columnwidth]{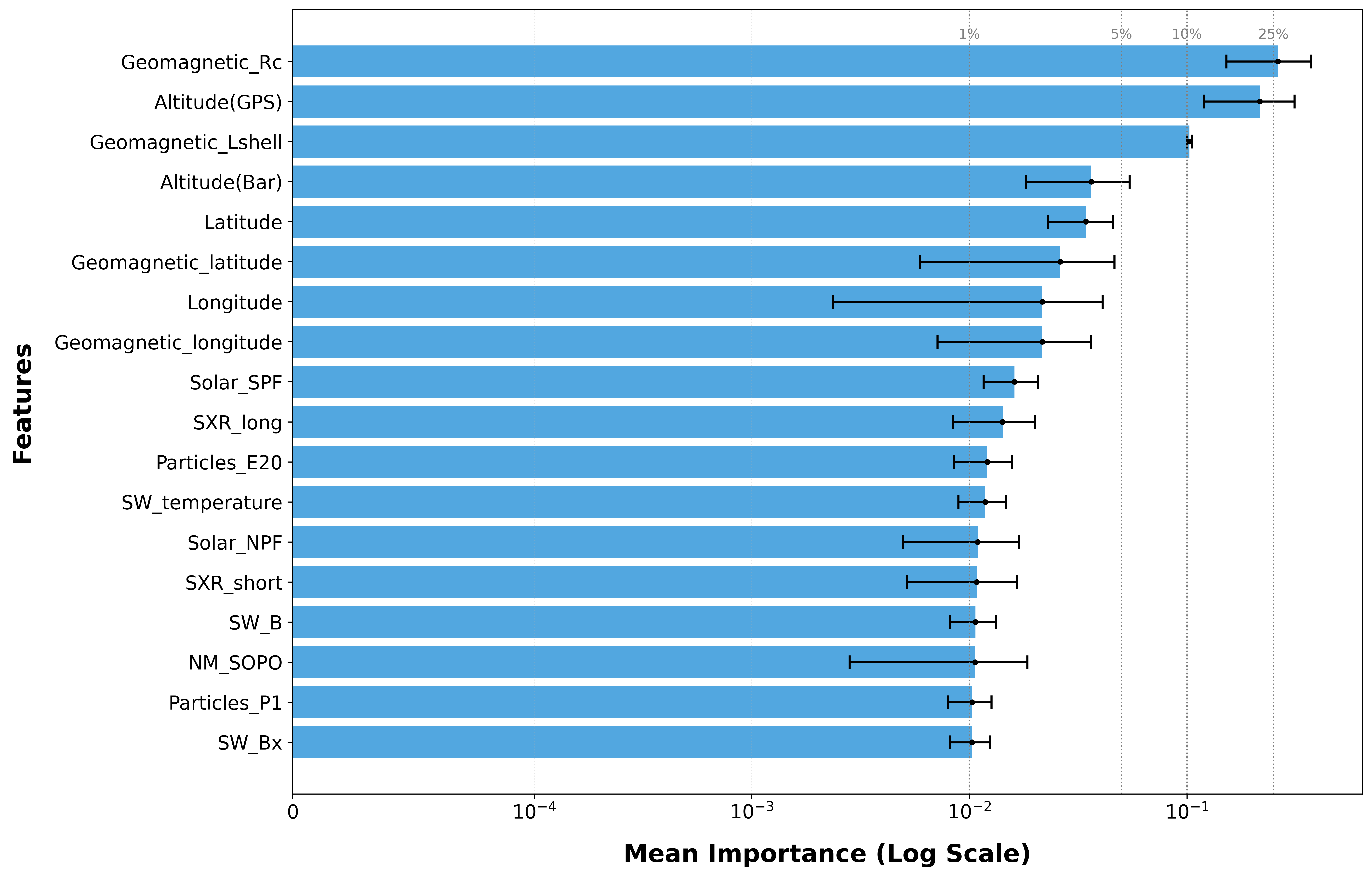}
        \caption{Mean feature importance according to the Random Forest model. }
        \label{fig:rffeatureimp}
    \end{figure*}

\begin{figure*}
        \centering
        \includegraphics[width=1.3\columnwidth]{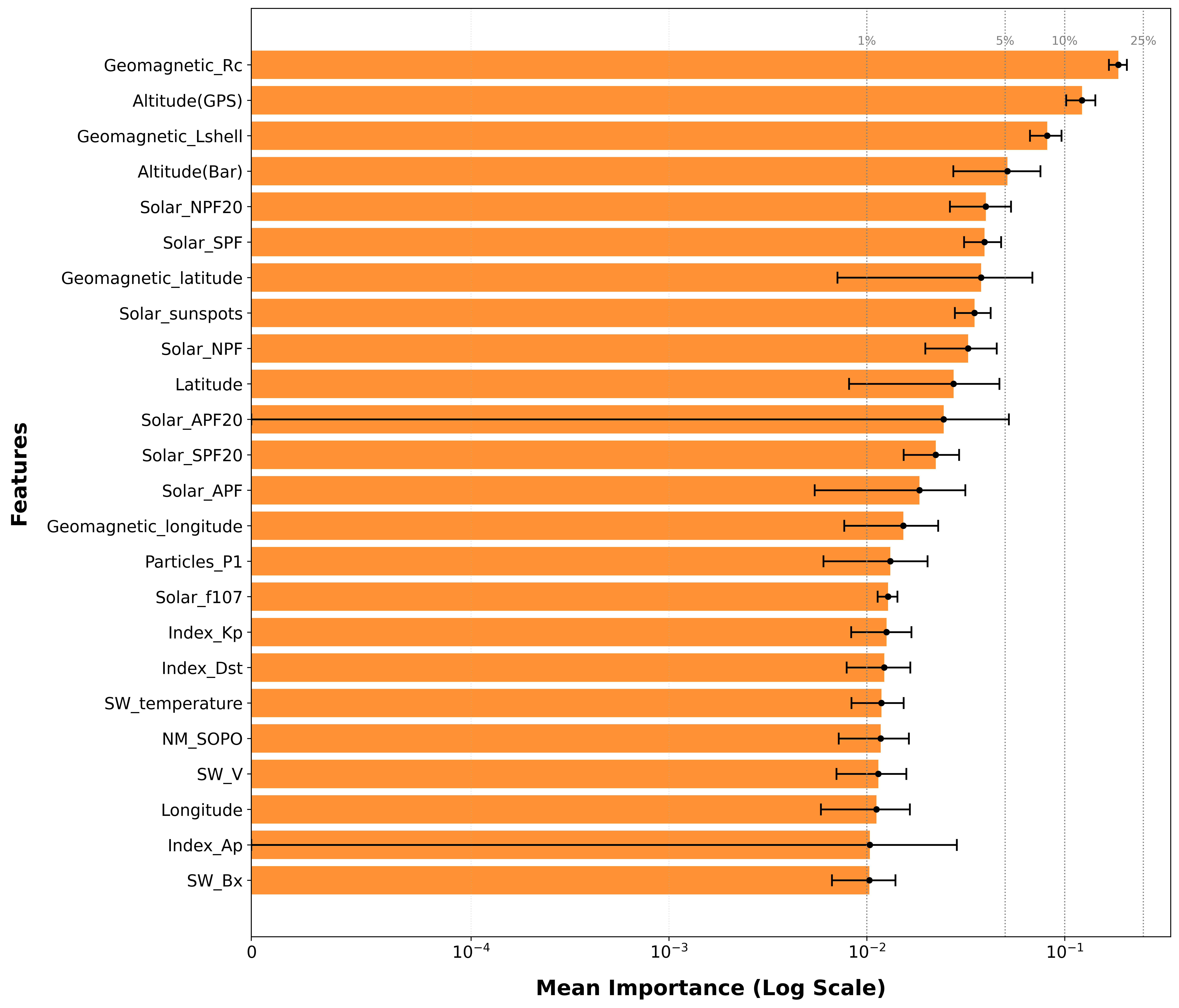}
        \caption{Mean feature importance according to the XGBoost model. The vertical lines show 1\%, 5\%, 10\%, and 25\% importance.}
        \label{fig:xgbfeatureimp}
    \end{figure*}

Figures~\ref{fig:rffeatureimp}~and~\ref{fig:xgbfeatureimp} display the mean feature importance scores \(\overline{I}_j\) obtained from the Random Forest and XGBoost model’s \texttt{feature\_importances\_} attribute respectively. This attribute measures the contribution of each feature \(j\) to the model’s predictive power—total variance reduction across trees for Random Forest, and gain (i.e., improvement in regression loss) for XGBoost. Importance values are averaged over all training splits, with only features exceeding 1\% importance shown. Each horizontal bar corresponds to a feature.

The x-axis uses a symmetric logarithmic scale (\texttt{symlog}) to represent a wide range of importance values, including very small ones. Vertical dotted lines at 1\%, 5\%, 10\%, and 25\% serve as reference points.

\textbf{Error bars explanation:}  
To visualize uncertainty in the importance estimates, asymmetric confidence intervals (CI) are used. For each feature \(j\), the CI is defined as
\[
\text{CI}_j = \overline{I}_j \pm
\begin{bmatrix}
\min(\sigma_j,\, \overline{I}_j) \\
\sigma_j
\end{bmatrix},
\]
where \(\overline{I}_j\) is the mean importance and \(\sigma_j\) is the standard deviation across model runs or splits. The lower bound is clipped at zero to avoid impossible negative values, allowing meaningful interpretation on the log scale. Black dots mark the exact mean importance values for clarity.

\subsection{Factors important for radiation} 

The feature importances combined from three considered models (LASSO regression, Random Forest, and XGBoost) can serve as a tool to understand the influence of different Geospace environment factors on the effective dose rates measured. In this subsection, we provide a qualitative assessment of the feature importance analysis and the corresponding physics implications.

The feature importance analysis for all three models indicates that geomagnetic cutoff rigidity and the GPS altitude represent the two most important features. This is understandable, as both quantify the energetic particle access to the Earth's atmosphere and, therefore, the flux of secondary cosmic ray particles. In particular, geomagnetic cutoff rigidities indicate the minimum momentum per charge for the particle to access the region, which is near-zero at the geomagnetic poles up to $\sim$15\,GV for some near-equatorial regions \cite{smart2009cutoff}. The energetic protons that can access the atmosphere start interacting with the atmospheric particles and produce secondary cosmic ray particles. Some of them, like muons, are born near the Regener-Pfotzer maximum height \cite{sarkar2017regenerpfotzer} (typically above civil aviation altitudes) and decay while propagating down to the ground. Therefore, higher heights would generally correspond to a higher dose rate \cite{shea2000dosewithheight}.

Other location-based features (such as geographic and geomagnetic longitudes and latitudes, barometric heights, or L-shells that also describe the magnetic environment) are also typically recognized as important. These especially stand out for Random Forest; although one has to note that this is likely due to (1) the limited number of features propagating to each tree \cite{breiman2001randomforest}, and (2) the degeneracy in how to describe the location of the measurement. Another interesting moment is that the GPS altitude was recognized to be several times more important than the barometric altitude (for Random Forest and XGBoost models), with the latter not being picked up as important for LASSO. Given that the physics-based radiation environment forecasting models typically operate on the barometric altitude grid \cite{mertens2025NAIRAS}, this potentially can indicate limitations of physics-based models.

Solar polar magnetic fields (quantified by the features of south polar field strength `Solar\_SPF', north polar field strength `Solar\_NPF', average polar field strength `Solar\_APF', and the corresponding 20\,nHz-filtered quantities with `20' postfix in the names) generally appear the next important features overall, and the most importance features describing the Geospace environment, for nowcasting of atmospheric radiation. Specifically, `Solar\_SPF' and `Solar\_NPF' features have the relative importance of 1\% or more for all three considered ML models. As the LASSO analysis indicates (see Figure~\ref{fig:lassofeatureimp}), both features contribute positively, i.e., the radiation environment is enhanced during the enhanced strengths of the polar fields. This behavior is expected: the polar magnetic fields are stronger during the solar minimum, when the GCR background is higher.

The other features that quantify the global solar activity and the GCR background are the data from neutron monitors, the 10.7\,cm radio flux, and the daily sunspot number. While none of them appear as important consistently across the models, it is still essential to highlight some peculiarities. Specifically, the neutron monitor data from South Pole Station (`NM\_SOPO') and Thule Station (`NM\_THUL') are recognized as important. These two stations are located at a lower geomagnetic cutoff rigidity ($R_c$=0.10\,GV and $R_c$=0.30\,GV, correspondingly) than the other two stations in the data set, Newark ($R_c$=2.40\,GV) and Oulu ($R_c$=0.81\,GV), and therefore are more susceptible to lower-energy GCRs. The 10.7\,cm radio flux and the daily sunspot numbers appear important only for the XGBoost model (see Figure~{\ref{fig:xgbfeatureimp}).

There are features that are supposed to have a less intuitive impact on the radiation environment but yet appear to be important. For example, all three models indicate that the properties of the solar wind are important for radiation forecasting. Every model has at least three solar wind properties at L1 recognized as important, with the solar wind temperature (`SW\_temperature') appearing systematically. Other properties include solar wind velocity or magnetic field components, but appear less consistently overall. One of the possible explanations for the dependence of the radiation field on the solar wind properties at L1 is the impact of the coronal mass ejections (CMEs). CME structures are denser than the surrounding solar wind. In addition, when arriving on Earth, CMEs can provide an additional magnetic shielding from the GCRs known as the Forbush decrease \cite{Mubashir2023Forbush}. Another possible reason is the general variation of the solar wind properties with the solar activity cycle \cite{Schwadron2011ApJ...739....9S}, which modulates the GCR precipitation as well.

Other properties indicated as important by different models are the energetic protons of different energies ($>$1\,MeV, `Particles\_P1'; $>$50\,MeV, `Particles\_P50'; $>$100\,MeV, `Particles\_P100') and energetic electrons ($>$2\,MeV, `Particles\_E20'), geomagnetic indexes (Kp, `Index\_Kp'; Dst, `Index\_Dst'; Ap `Index\_Ap'), and fluxes of soft X-ray radiation (1-8$\AA$, `SXR\_long'; 0.5-4$\AA$, `SXR\_short'). However, since these features do not appear consistently for each of the considered ML models and vary with the solar cycle progression in general, it is difficult to say whether their importance is beyond the general solar cycle trend.

\section{Conclusions} 
In this work, we have tested several machine learning techniques (Random Forest, XGBoost, and LASSO regression) for the problem of nowcasting of the atmospheric radiation at aviation altitudes. The results for all the classic ML models show promise in the nowcasting problem of aviation radiation, comparable to and even better than the physics-based model considered (NAIRAS-v3). This suggests the possibility of developing a scalable, efficient, and global ML-driven operational model. The rich feature space helped to analyze, compare, and direct some possible physics-based improvements, which can even be expanded in the future. A more sophisticated model, like deep neural networks for the static problem or RNN-type architectures for the time series-driven predictions, can be applied and fine-tuned for better output in the future. Our future research incorporates this, including the analysis of time-series-based data for the Geospace parameters to predict the aviation radiation within the safe time window required for the operational model.

\section*{Acknowledgments} 
This work was supported by the NSF FDSS grant 1936361, NASA LWS grant 80NSSC24K1111, and NASA HITS grant 80NSSC22K1561.

\balance
\bibliographystyle{IEEEtran}
\bibliography{references}

\begin{thebibliography}{10}
\providecommand{\url}[1]{#1}
\csname url@samestyle\endcsname
\providecommand{\newblock}{\relax}
\providecommand{\bibinfo}[2]{#2}
\providecommand{\BIBentrySTDinterwordspacing}{\spaceskip=0pt\relax}
\providecommand{\BIBentryALTinterwordstretchfactor}{4}
\providecommand{\BIBentryALTinterwordspacing}{\spaceskip=\fontdimen2\font plus
\BIBentryALTinterwordstretchfactor\fontdimen3\font minus \fontdimen4\font\relax}
\providecommand{\BIBforeignlanguage}[2]{{%
\expandafter\ifx\csname l@#1\endcsname\relax
\typeout{** WARNING: IEEEtran.bst: No hyphenation pattern has been}%
\typeout{** loaded for the language `#1'. Using the pattern for}%
\typeout{** the default language instead.}%
\else
\language=\csname l@#1\endcsname
\fi
#2}}
\providecommand{\BIBdecl}{\relax}
\BIBdecl

\bibitem{url-rdpmldataset}
\BIBentryALTinterwordspacing
V.~M. Sadykov, Z.~M. Watkins, D.~Kempton, W.~Jones, S.~K~C, G.~T. Goodwin, X.~He, W.~K. Tobiska, I.~Kitiashvili, C.~Mertens, S.~Ranjan, D.~G. Deardorff, R.~Spaulding, and S.~L. Pokrandt, ``{ML-Ready Data Set for Aviation Radiation Nowcasting},'' 2025. [Online]. Available: \url{https://dmlab.cs.gsu.edu/rdp/ml-dataset.html}
\BIBentrySTDinterwordspacing

\bibitem{friedberg2003aircrews}
W.~Friedberg, K.~Copeland \emph{et~al.}, ``What aircrews should know about their occupational exposure to ionizing radiation,'' United States. Department of Transportation. Federal Aviation Administration~…, Tech. Rep., 2003.

\bibitem{Tobiska2016ARMAS}
W.~K. {Tobiska}, D.~{Bouwer}, D.~{Smart}, M.~{Shea}, J.~{Bailey}, L.~{Didkovsky}, K.~{Judge}, H.~{Garrett}, W.~{Atwell}, B.~{Gersey}, R.~{Wilkins}, D.~{Rice}, R.~{Schunk}, D.~{Bell}, C.~{Mertens}, X.~{Xu}, M.~{Wiltberger}, S.~{Wiley}, E.~{Teets}, B.~{Jones}, S.~{Hong}, and K.~{Yoon}, ``{Global real-time dose measurements using the Automated Radiation Measurements for Aerospace Safety (ARMAS) system},'' \emph{Space Weather}, vol.~14, no.~11, pp. 1053--1080, Nov. 2016.

\bibitem{straume2016ground}
T.~Straume, C.~Mertens, T.~Lusby, B.~Gersey, W.~Tobiska, R.~Norman, G.~Gronoff, and A.~Hands, ``Ground-based evaluation of dosimeters for nasa high-altitude balloon flight,'' \emph{Space Weather}, vol.~14, no.~11, pp. 1011--1025, 2016.

\bibitem{Tobiska2015ARMAS}
W.~K. {Tobiska}, W.~{Atwell}, P.~{Beck}, E.~{Benton}, K.~{Copeland}, C.~{Dyer}, B.~{Gersey}, I.~{Getley}, A.~{Hands}, M.~{Holland}, S.~{Hong}, J.~{Hwang}, B.~{Jones}, K.~{Malone}, M.~M. {Meier}, C.~{Mertens}, T.~{Phillips}, K.~{Ryden}, N.~{Schwadron}, S.~A. {Wender}, R.~{Wilkins}, and M.~A. {Xapsos}, ``{Advances in Atmospheric Radiation Measurements and Modeling Needed to Improve Air Safety},'' \emph{Space Weather}, vol.~13, no.~4, pp. 202--210, Apr. 2015.

\bibitem{Tobiska2018ARMAS}
W.~K. {Tobiska}, L.~{Didkovsky}, K.~{Judge}, S.~{Weiman}, D.~{Bouwer}, J.~{Bailey}, B.~{Atwell}, M.~{Maskrey}, C.~{Mertens}, Y.~{Zheng}, M.~{Shea}, D.~{Smart}, B.~{Gersey}, R.~{Wilkins}, D.~{Bell}, L.~{Gardner}, and R.~{Fuschino}, ``{Analytical Representations for Characterizing the Global Aviation Radiation Environment Based on Model and Measurement Databases},'' \emph{Space Weather}, vol.~16, no.~10, pp. 1523--1538, Oct. 2018.

\bibitem{valentin2007international}
J.~Valentin, ``International commission on radiological protection,'' \emph{ICRP Publication}, vol. 103, 2007.

\bibitem{sadykov2025operational}
V.~Sadykov, P.~Martens, D.~Kempton, R.~Angryk, B.~Aydin, J.~Hamilton, G.~Goodwin, A.~Ali, R.~Syeda, I.~Kitiashvili \emph{et~al.}, ``Operational and exploration requirements and research capabilities for sep environment monitoring and forecasting,'' \emph{arXiv preprint arXiv:2505.10390}, 2025.

\bibitem{tobiska2019armas}
K.~Tobiska, L.~Didkovsky, K.~Judge, D.~Bouwer, S.~Wieman, B.~Gersey, B.~Atwell, and R.~Wilkins, ``Armas flight system for operational aerospace radiation measurements.''\hskip 1em plus 0.5em minus 0.4em\relax 49th International Conference on Environmental Systems, 2019.

\bibitem{Sadykov2021RDP}
V.~M. {Sadykov}, I.~N. {Kitiashvili}, W.~K. {Tobiska}, and M.~{Guhathakurta}, ``{Radiation Data Portal: Integration of Radiation Measurements at the Aviation Altitudes and Solar-Terrestrial Environment Observations},'' \emph{Space Weather}, vol.~19, no.~1, p. e2020SW002653, Jan. 2021.

\bibitem{mertens2009development}
C.~Mertens, W.~K. Tobiska, D.~Bouwer, B.~Kress, M.~Wiltberger, S.~Solomon, and J.~Murray, ``Development of nowcast of atmospheric ionizing radiation for aviation safety (nairas) model,'' in \emph{1st aiaa atmospheric and space environments conference}, 2009, p. 3633.

\bibitem{Mertens2013}
C.~J. {Mertens}, M.~M. {Meier}, S.~{Brown}, R.~B. {Norman}, and X.~{Xu}, ``{NAIRAS aircraft radiation model development, dose climatology, and initial validation},'' \emph{Space Weather}, vol.~11, no.~10, pp. 603--635, Oct. 2013.

\bibitem{Mertens2023}
C.~J. {Mertens}, G.~P. {Gronoff}, Y.~{Zheng}, M.~{Petrenko}, J.~{Buhler}, D.~{Phoenix}, E.~{Willis}, I.~{Jun}, and J.~{Minow}, ``{NAIRAS Model Run-On-Request Service at CCMC},'' \emph{Space Weather}, vol.~21, no.~5, p. e2023SW003473, May 2023.

\bibitem{Mertens2024}
C.~J. Mertens, G.~P. Gronoff, Y.~Zheng, J.~Buhler, E.~Willis, M.~Petrenko, D.~Phoenix, I.~Jun, and J.~Minow, ``Nairas atmospheric and space radiation environment model,'' \emph{IEEE Transactions on Nuclear Science}, vol.~71, no.~4, pp. 618--625, 2024.

\bibitem{mertens2025NAIRAS}
C.~J. {Mertens}, G.~P. {Gronoff}, and D.~B. {Phoenix}, ``{NAIRAS Version 3 Atmospheric Ionizing Radiation Validation: Comparisons to RaD-X Measurements},'' \emph{Space Weather}, vol.~23, no.~4, p. e2024SW004296, Apr. 2025.

\bibitem{copeland2021cari7}
\BIBentryALTinterwordspacing
K.~Copeland, ``\BIBforeignlanguage{English}{Cari-7 documentation: Radiation transport in the atmosphere},'' United States. Department of Transportation. Federal Aviation Administration. Office of Aviation. Civil Aerospace Medical Institute, Tech Report DOT/FAA/AM-21/05, Mar. 2021. [Online]. Available: \url{https://rosap.ntl.bts.gov/view/dot/57224}
\BIBentrySTDinterwordspacing

\bibitem{phoenix2024characterization}
D.~B. Phoenix, C.~J. Mertens, G.~P. Gronoff, and K.~Tobiska, ``Characterization of radiation exposure at aviation flight altitudes using the nowcast of aerospace ionizing radiation system (nairas),'' \emph{Space Weather}, vol.~22, no.~4, p. e2024SW003869, 2024.

\bibitem{url-ccmc-nairasv3}
\BIBentryALTinterwordspacing
C.~Mertens and Y.~Zheng, ``{NAIRAS},'' 2025. [Online]. Available: \url{https://ccmc.gsfc.nasa.gov/models/NAIRAS~3.0/}
\BIBentrySTDinterwordspacing

\bibitem{url-rdpdmlab}
\BIBentryALTinterwordspacing
D.~Kempton, V.~Sadykov, and I.~Kitiashvili, ``{Radiation Data Portal},'' 2025. [Online]. Available: \url{dmlab.cs.gsu.edu/rdp}
\BIBentrySTDinterwordspacing

\bibitem{sadykov2025mlradiation}
V.~M. Sadykov, Z.~M. Watkins, D.~Kempton, W.~Jones, S.~K~C, G.~T. Goodwin, X.~He, W.~K. Tobiska, I.~Kitiashvili, C.~Mertens, S.~Ranjan, D.~G. Deardorff, R.~Spaulding, and S.~L. Pokrandt, ``Machine learning-ready data sets for the analysis and nowcasting of atmospheric radiation at aviation altitudes,'' \emph{Space Weather}, 2025, under review.

\bibitem{tibshirani1996regression}
R.~Tibshirani, ``Regression shrinkage and selection via the lasso,'' \emph{Journal of the Royal Statistical Society Series B: Statistical Methodology}, vol.~58, no.~1, pp. 267--288, 1996.

\bibitem{tibshirani2011regression}
------, ``Regression shrinkage and selection via the lasso: a retrospective,'' \emph{Journal of the Royal Statistical Society Series B: Statistical Methodology}, vol.~73, no.~3, pp. 273--282, 2011.

\bibitem{breiman2001random}
L.~Breiman, ``Random forests,'' \emph{Machine learning}, vol.~45, no.~1, pp. 5--32, 2001.

\bibitem{chen2016xgboost}
T.~Chen and C.~Guestrin, ``Xgboost: A scalable tree boosting system,'' in \emph{Proceedings of the 22nd acm sigkdd international conference on knowledge discovery and data mining}, 2016, pp. 785--794.

\bibitem{ali2024predicting}
A.~Ali, V.~Sadykov, A.~Kosovichev, I.~N. Kitiashvili, V.~Oria, G.~M. Nita, E.~Illarionov, P.~M. O’Keefe, F.~Francis, C.-J. Chong \emph{et~al.}, ``Predicting solar proton events of solar cycles 22--24 using goes proton and soft-x-ray flux features,'' \emph{The Astrophysical Journal Supplement Series}, vol. 270, no.~1, p.~15, 2024.

\bibitem{Sadykov2025ionkinetic}
V.~M. {Sadykov}, L.~{Ofman}, S.~A. {Boardsen}, {Yogesh}, P.~{Mostafavi}, L.~K. {Jian}, K.~{Klein}, and M.~{Martinovi{\'c}}, ``{Identification of Ion-kinetic Instabilities in Hybrid-PIC Simulations of Solar Wind Plasma with Machine Learning},'' \emph{The Astrophysical Journal Supplement Series}, vol. 279, no.~1, p.~28, Jul. 2025.

\bibitem{pedregosa2011scikit}
F.~Pedregosa, G.~Varoquaux, A.~Gramfort, V.~Michel, B.~Thirion, O.~Grisel, M.~Blondel, P.~Prettenhofer, R.~Weiss, V.~Dubourg \emph{et~al.}, ``Scikit-learn: Machine learning in python,'' \emph{the Journal of machine Learning research}, vol.~12, pp. 2825--2830, 2011.

\bibitem{smart2009cutoff}
D.~F. {Smart} and M.~A. {Shea}, ``{Fifty years of progress in geomagnetic cutoff rigidity determinations},'' \emph{Advances in Space Research}, vol.~44, no.~10, pp. 1107--1123, Nov. 2009.

\bibitem{sarkar2017regenerpfotzer}
R.~{Sarkar}, S.~K. {Chakrabarti}, P.~S. {Pal}, D.~{Bhowmick}, and A.~{Bhattacharya}, ``{Measurement of secondary cosmic ray intensity at Regener-Pfotzer height using low-cost weather balloons and its correlation with solar activity},'' \emph{Advances in Space Research}, vol.~60, no.~5, pp. 991--998, Sep. 2017.

\bibitem{shea2000dosewithheight}
M.~A. {Shea} and D.~F. {Smart}, ``{Cosmic ray Implications for Human Health},'' \emph{Space Science Reviews}, vol.~93, pp. 187--205, Jul. 2000.

\bibitem{breiman2001randomforest}
L.~{Breiman}, ``{Random Forests.}'' \emph{Machine Learning}, vol.~45, pp. 5--32, Jan. 2001.

\bibitem{Mubashir2023Forbush}
A.~{Mubashir}, A.~{Ashok}, A.~G. {Bourgeois}, Y.~T. {Chien}, M.~{Connors}, E.~{Potdevin}, X.~{He}, P.~{Martens}, A.~{Mikler}, A.~G.~U. {Perera}, V.~{Sadykov}, M.~{Sarsour}, D.~{Sharma}, and C.~{Tiwari}, ``{Muon Flux Variations Measured by Low-Cost Portable Cosmic Ray Detectors and Their Correlation With Space Weather Activity},'' \emph{Journal of Geophysical Research (Space Physics)}, vol. 128, no.~12, p. e2023JA031943, Dec. 2023.

\bibitem{Schwadron2011ApJ...739....9S}
N.~A. {Schwadron}, C.~W. {Smith}, H.~E. {Spence}, J.~C. {Kasper}, K.~{Korreck}, M.~L. {Stevens}, B.~A. {Maruca}, K.~K. {Kiefer}, S.~T. {Lepri}, and D.~{McComas}, ``{Coronal Electron Temperature from the Solar Wind Scaling Law throughout the Space Age},'' \emph{The Astrophysical Journal}, vol. 739, no.~1, p.~9, Sep. 2011.

\end{thebibliography}

\end{document}